\documentclass[a4paper]{article}

\usepackage[T1]{fontenc}
\usepackage[utf8x]{inputenc}
\usepackage[english]{babel}

\usepackage[colorlinks=true, allcolors=blue]{hyperref}

\newcommand{\email}[1]{\href{mailto:#1}{\tt{\nolinkurl{#1}}}}
\newcommand{\orcid}[1]{ORCID: \href{https://orcid.org/#1}{\tt{\nolinkurl{#1}}}}

\usepackage[sfdefault,lf]{carlito}
\usepackage[parfill]{parskip}

\usepackage{fancyhdr}
\usepackage{authblk}
\usepackage{mathtools}

\usepackage[a4paper,top=3cm,bottom=2cm,left=3cm,right=3cm,marginparwidth=1.75cm]{geometry}

\usepackage{amsmath}
\usepackage{graphicx}
\usepackage{booktabs}
\usepackage{graphbox}

\title{Network structure and disease risk for an endemic infectious disease}
\author[1,*]{Jose L. Herrera-Diestra}
\author[2]{Michael Tildesley}
\author[1,3]{Katriona Shea}
\author[1,3]{Matthew Ferrari}
\affil[1]{Department of Biology, The Pennsylvania State University, University Park, PA, USA.}
\affil[2]{Zeeman Institute for Systems Biology and Infectious Disease Epidemiology Research, Mathematics Institute and School of Life Sciences, University of Warwick, Coventry CV47AL, UK.}
\affil[3]{Center for Infectious Disease Dynamics, Pennsylvania State University, University Park, PA, USA.}
\affil[*]{Corresponding author: \email{diestra@austin.utexas.edu}}
\date{}

\begin{document}
\maketitle
\thispagestyle{fancy}

\begin{abstract}
The structure of contact networks affects the likelihood of disease spread at the population scale and the risk of infection at any given node. Though this has been well characterized for both theoretical and empirical networks for the spread of epidemics on completely susceptible networks, the long-term impact of network structure on risk of infection with an endemic pathogen, where nodes can be infected more than once, has been less well characterized. Here, we analyze detailed records of the transportation of cattle between farms in Turkey to characterize the global and local attributes of the shipments network between 2007-2012. We build an aggregated static directed - weighted network, using the information about source-destination of shipments and the frequency of shipments between farms as weights. We then study the correlation between network properties and the likelihood of infection with, or exposure to, foot-and-mouth (FMD) disease over the same time period using recorded outbreaks. The shipments network shows properties of "small-worldness" in that the shortest path length is small (similar to that of a random-equivalent networks ensemble), combined with a large clustering coefficient (much larger than on a random network). The degree distribution is scale-free over the intermediate range of degrees, but has an exponential cut-off for high degrees. Further, high degree nodes disproportionately have more frequent shipments on average. In addition, the shipments network illustrates strong modular structure, where farms are more preferentially attached to other geographically proximate farms than to distant farms; and lack of assortativity, i.e. farms connecting other farms regardless of their connections. This combination of features has not been previously observed in other reported networks of shipments. Locally, the shipments network shows signs of spatial constraints, with relatively few long-distance connections, and a strong similarity to other spatially constrained networks. We find that farms that were either infected or at high risk of infection with FMD (within one link from an infected farm) had higher values of centrality (eigenvector centrality, betweenness centrality, degree, coreness); farms that were never less than 2 links from an infected farm had disproportionately low centrality. 
However, the correlation of the rankings of farms shows that central farms (high eigenvector centrality) are not necessarily those with more connections to/from it (in/out degree). Several farms with influential connections (high eigenvector centrality) serve as \emph{bridges} of densely connected farms (high betweenness centrality); i.e., are connections between modules.
These results suggest that to detect FMD spread, surveillance efforts could be focused preferentially on farms with centralities greater than the mean.
\end{abstract}

\section{Introduction}

The contact structure of a population, in particular heterogeneity in the number or rate of potential contacts between individuals, is an important predictor of infectious disease transmission \cite{albert2000,watts1998,newman2003,eubank2004,meyers2005,CHOWELL201666}. 
The theoretical relationship between contact structure and disease transmission dynamics has been illustrated using a variety of epidemiologically relevant contact databases \cite{kimLee2018,herrera2016,bai2017,ames2011}. 
Theory suggests that measures of contact structure are predictive of infection risk and the potential to transmit to others. Different global (eigenvector centrality, degree, coreness, betweenness) and local (random acquaintance \cite{cohen2003}) network measures have been used to propose surveillance and vaccination strategies in static \cite{herrera2016,colman2019,schirdewahn2021} and temporally varying \cite{bai2017,holme2012} networks. Thus, \emph{a priori} network characterization can be used to identify candidate sites for detecting outbreaks, either in static  \cite{albert2000,watts1998,meyers2005,Lloyd2001,keeling2005,newman2002}, temporal \cite{rocha_individual-based_2016,Karsai2017}, dynamic \cite{kao2006}  or adaptive networks \cite{gross_epidemic_2006}. Moreover, these measures and their correlations are predictive of epidemic spread and can facilitate rapid targeting of interventions once an outbreak starts \cite{schwartz2002,allard2020}.
The role of contact structure on the initial spread of infection in a naive population has been well characterized, including for the current pandemic of COVID-19 \cite{Thurner22684}. 
The relationship between contact structure and long-term infection risk for an endemic disease is less well characterized \cite{vazquezF2016,rothenberg2000,Eames13330,ghani2000,doherty2005,eames_monogamous_2004,pastor2001a}.
An endemic infection may experience the network topology differently from a novel outbreak \cite{Eames13330,rotherberg2009,sloot2008}, as prior infection (e.g. immunity) or interventions may alter risk and transmission of infection \cite{ferrari2006}.

Livestock transportation records provide a rich resource for describing characteristics of livestock movement networks, including source location, destination, date, and batch size. Such records have been analyzed to characterize networks of interactions \cite{machado2020,lentz2016,valerio2020,kiss2006,sterchi2019,rowland2007,danon2011,GORSICH201682}, and the potential consequences of network structure on the spreading of infection \cite{lentz2016,silk2017,MOHR20188}. However, the lack of reliable data about the infection of farms combined with detailed shipment data, has hindered our ability to build data driven models to appropriately assess the relationship between disease incidence and contact network structure in a single population, with few exceptions \cite{pozo}. Instead, the impact of the interrelation between transportation network topology and disease has only been explored through theoretical simulations \cite{schirdewahn2021,machado2020,valerio2020,kiss2006,sterchi2019,danon2011,silk2017,ortiz2006}, or through the reconstruction of who-infected-whom networks \cite{haydon2004,hayama2019}, which describe the network of realized transmission of a specific outbreak rather than the network of potential transmission. The development of reliable surveillance systems, both for learning about and for managing emerging or endemic diseases remains challenging. There are many unknown characteristics of transmission and control that hamper accurate decision making, including the connectivity between farms, the duration of  immunity following infection, the role of multiple serotypes circulating in the livestock population and the use (and efficacy) of vaccines for spreading diseases, such as foot-and-mouth disease. Foot-and-mouth disease (FMD) is a highly contagious, viral disease of cloven-hoofed species (such as cattle, sheep, and pigs). In Turkey, FMD was eliminated in the Thrace region in 2010, but remains endemic in Anatolian Turkey. There are 7 immunologically distinct serotypes of FMD; two serotypes, A and O, have been endemic in Turkey continuously. A third serotype, Asia-1, has been present intermittently and re-emerged in Turkey in 2011 after going unrecorded  since 2001.

Here we analyze in detail the livestock transportation records of cattle in Turkey. We use records of shipments of cattle between farms (source, destination, date, batch size) to create an aggregated directed - weighted network. We then characterize the global features of the network, as well as the local ``\emph{importance}'' of each node in the network. Additionally, we analyze the occurrence of 3718 outbreaks of FMD in Turkey between 2007 and 2012 relative to the network defined by the livestock movement records. Over this time period, we compare the relationship between network structure and the risk of infection with either of the two endemic serotypes (A and O) to the corresponding pattern in the re-emergent Asia-1 serotype. We first describe the structure of the Turkey cattle movement network and then describe how network structure correlates with FMD outbreak risk at the node-level. We show that while some metrics  are correlated with the occurrence of an outbreak, they are more strongly predictive of the absence of outbreaks. Finally, we discuss how these observations can be integrated to develop plausible network-based ``active surveillance'' strategies for the placement of sentinel sites/farms, which follows a purposeful and comprehensive search for evidence of disease in animal or human populations \cite{geering1999}.

\section{Methodology}

\subsection{Shipment network}

\subsubsection{The data}

The data on cattle shipments was provided by the Turkish Veterinary authorities, facilitated by the European Commission for foot-and-mouth disease (EuFMD), who granted the access to data from the TurkVet database. The highest resolution of cattle farming unit in Turkey is the holding, of which there are over 2.9 million. In each holding, birth, movement, and death data are recorded. The number of animals on these holdings can range from fewer than 5 to over 500. Since many of these holdings are small, the basic epidemiological unit for recording FMD outbreaks in Turkey is the epiunit (a village or a neighbourhood comprised of several holdings \cite{dawson2016}); there were 49850 epiunits in Turkey (Figure \ref{fig:01}{\bf a}) after aggregation. (A detailed description of the data is in Section 1 supplemental material). 

\subsubsection{Network construction}

We create the livestock movement (shipments) network, where each epiunit is represented by a node, and an edge is placed between two epiunits if there exists a shipment of cattle between them. In general, a graph $G$ can be defined as a pair $(V,E)$, where $V$ is a set of vertices, and $E$ is a set of edges between the vertices $E \subseteq {(u,v) | u, v \in V}$. A graph or network can be represented as an adjacency matrix $A$, defined as
\begin{equation}
    A =
    \begin{cases*}
      A_{ij}=w_{ij} & if $i$ is connected to $j$ \\
      0        & otherwise.
    \end{cases*}
\end{equation}

The network considered here is an aggregation of all shipments between epiunits, which creates a static - aggregated network (from here on called the \emph{shipments network}). The shipments network is directed (there is information about origin - destination of shipments) and weighted; the weights, $w_{ij}$, are calculated using the frequency of unique shipments, regardless of the number of animals in the shipment, between epiunits. 

\subsection{Network statistics}

Among the plethora of statistics used to characterize a complex network, we focus on those that have been used in previous studies and shown to be related to spreading dynamics; particularly epidemics \cite{kiss2006,danon2011,silk2017,Barrat3747,kitsak2010,Batagelj02,newmanBook,newmanSW2001}.

We calculated global measures, which evaluate the overall network properties (how nodes organize globally): density; average shortest path length ($L$); diameter ($D$); degree assortativity ($\rho$); giant strongly connected components ($GSCC$); giant weakly connected components ($GWCC$); largest eigenvalue of the adjacency matrix ($\lambda_1$); reciprocity; the global clustering coefficient ($C$); and modularity ($Q$) using the Infomap community detection algorith \cite{infomap2008}. 
It is not possible to assess if a given global network measure is ``high'' or ``low'' by itself; e.g. "high" or "low" clustering must be compared to the likelihood of observing these measures in appropriate null models of random networks. In order to determine if the structure and characteristics of the shipments network emerge as a consequence of its innate dynamics, we compare the observed metrics with a set of 100 random networks created using the \emph{local weight reshuffling} method \cite{opsahl2008}, which is a randomisation procedure for directed networks that preserves the strength (sum of the weights of all connections) of a node by reshuffling weights locally for each node across its outgoing ties, in addition to degree and strength distribution. We then present a Z-score for each metric, which is calculated as  Z-score $= (x-\mu)/\sigma$, where $x$ is the calculated measure in the actual network, and $\mu$ and $\sigma$ are the average value and the standard deviation of the given measure calculated in the random ensemble, respectively. Thus, the Z-score reflects the likelihood of the observed value of the metric arising on a random equivalent network.

We calculated local (node level) measures, which evaluate the position of each epiunit in the network: degree ($k_i^{in/out}$), and its corresponding distribution; strength ($s_i^{in/out}$), and its corresponding distribution; local clustering coefficient ($c_i^w$); $k$-coreness ($kC_i^{in/out}$); eigenvector centrality ($ec(i)$) and relative betweenness centrality ($B_i$). In each measure $i$ refers to epiunits and in/out refers to the direction of the edges used to calculate the measure. Due to the large size of the shipments network and computational limitations, the calculation of $L$ and $D$ (lower bound) were performed in the directed - unweighted version of the network; while $C$ and $Q$ (upper bound) were calculated on the undirected - weighted version of the network.

To compare the structure of the livestock shipments network to conventional, well-studied, networks, we built correlation matrices of local measures for 9 published networks from different contexts: neuronal, social and transportation (the description of these networks is detailed in the supplemental material, Section 4). We then calculate the Spearman correlation between the pair-wise correlations of local measures in the Turkey shipments network and each of the 9 published networks. All calculations were performed in R, and those related to complex networks statistics with the R package igraph \cite{igraphR2006}. A detailed description about the definitions of the local and global measures can be found in Section 2 of the supplemental material.

In addition to these measures, we introduce a simple measure that compares the balance of in and out shipments for each epiunit. We define the ``\emph{transmission flux}'' ($\phi$) of each epiunit as
\begin{equation}
\phi_X=\frac{X_{out}-X_{in}}{X_{in}+X_{out}},
\end{equation}
where $X$ represents the degree, strength, or coreness of epiunits (measures that were calculated using in and out modes). Note that $\phi_X\in[-1,1]$. When an epiunit has $\phi_X=1$, $X_{in}=0$ and $X_{out}\neq0$, it is a net ``source'' of shipments. A value of $\phi_X=-1$, $X_{out}=0$ and $X_{in}\neq0$, identifies net ``sink'' epiunits. An epiunit with  $\phi_X=0$ and $X_{in},X_{out}\neq0$, sends as many shipments as it receives.
Our transmission flux ($\phi_X$) discriminates between epiunits according to their vulnerability to get infected from a spreading disease (sinks) and their ability to transmit infection (sources) \cite{allard2020}.

\begin{figure}[!ht]
\centering
\includegraphics[width=\linewidth]{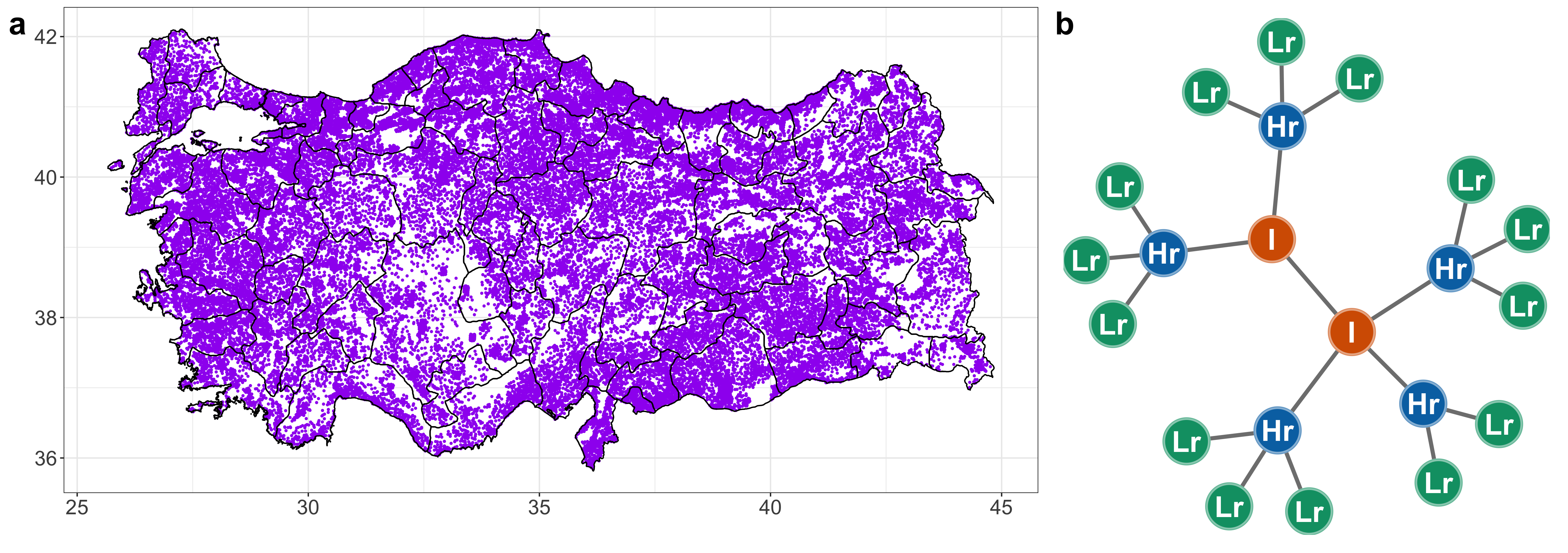}
\caption{\label{fig:01} {\bf a.} Location of epiunits in Turkey. Each purple dot represents the location of an epiunit in the map. {\bf b.} Schematic representation for the definition classes of epiunits: Infected (red); high-risk (blue) and low-risk (green).}
\end{figure}

\subsection{FMD spreading in Turkey}

In addition to the data of shipments between epiunits, $6112$ outbreaks of FMD (with identified serotypes) were reported to TurkVet between January 2001 and July 2012. The data include epiunit locations and dates of outbreaks. From these data, we select the period of time which overlaps with the shipment data (from 2007 to 2012), which included $3718$ outbreaks.

For each epiunit in the shipments network we assign a state label, which depends on their ``distance'' from outbreaks as follows (Figure \ref{fig:01}{\bf b}):
\begin{enumerate}
    \item \emph{Infected epiunits (I)} (red); epiunits that experienced at least one outbreak of FMD,
    \item \emph{High-risk epiunits (Hr)} (blue); epiunits that were directly connected (through at least one shipment) to an epiunit that experienced a FMD outbreak.
    \item \emph{Low-risk epiunits (Lr)} (green); epiunits that were at least at distance two (two degrees of separation) from an infected epiunit.
\end{enumerate}

\section{Results}

\subsection{Characterization of the shipments network}

The shipments network is a directed-weighted network which consists of $N=49,580$ nodes (epiunits) which are connected with $E=4,746,035$ edges. Though the density of edges is low, $\approx0.2\%$ of all the possible edges that could exist in the network, the shipments network has very strong connectivity ($\lambda_1$, Z-score$\approx200$). The shipments network has a giant strongly connected component (GSCC)  that spans to $97\%$  of epiunits of the network; the giant weakly connected component (GWCC) spans $100\%$  of epiunits (Table \ref{tab:globalNet}).

\begin{table}[!ht]
\centering
 \caption{Network measures for the shipments network (\text{*}Calculated using the directed-unweighted version of the network. \text{**}Calculated using the undirected-weighted version of the network)}
 \label{tab:globalNet}
 \begin{tabular}{| c | c | c |} 
 \hline
 Measure & Value & Z-score \\ [0.5ex] 
 \hline\hline
 Density & $0.002$ & - \\
 Shortest path length ($L$) & $2.86$\text{*} & $0.01$\\ 
 Diameter ($D$) & $20$\text{*} & $0.32$ \\
 Assortativity ($\rho$) & $-0.04$ & $-4.35$ \\ 
 GSCC & 97\% & - \\
 GWCC & 100\% & - \\
 Largest eigenvalue ($\lambda_1$) & $3280.40$ & $195.03$ \\
 Reciprocity & $0.34$ & $-494.61$ \\
 Clustering coeff. ($C$)& $0.51\text{**}$ & $23.69$\\
 Modularity(Q) & $0.67$\text{**} & $1540.97$ \\
\hline
 \end{tabular}
\end{table}

 The shipments network shows strong evidence of small-worldness; the clustering coefficient ($C$) is high and much larger than its random counterpart (Z-score$\approx24$), and an average shortest path that satisfies the relation $L< \log(N)$ \cite{watts1998} (Table \ref{tab:globalNet}). The diameter of the network ($20$), which similar to the random equivalent counterpart, implies that information (i.e., a disease) could spread throughout the weighted edges of the shipments network in a few transmission events (Table \ref{tab:globalNet}).

The shipments network shows a slight tendency to a disassortative mixing, i.e., epiunits tend to connect to other epiunits with different degrees ($\rho<0$); however, since $\rho\approx0$, patterns of connectivity between epiunits may be similar to a proportionate mixing (Table \ref{tab:globalNet}). Only $\approx1/3$ of links are reciprocal, which is a considerably smaller value than for random equivalent networks (Z-score$\approx-500$). The shipments network has a strong modular structure ($Q=0.67$); which is very different (higher) from the random equivalent ensemble, as shown by the large Z-score (Table \ref{tab:globalNet}). There were $n\approx110$, where $n$ is the number of modules with more than 10 epiunits.

\begin{figure}[!ht]
\centering
\includegraphics[width=\textwidth]{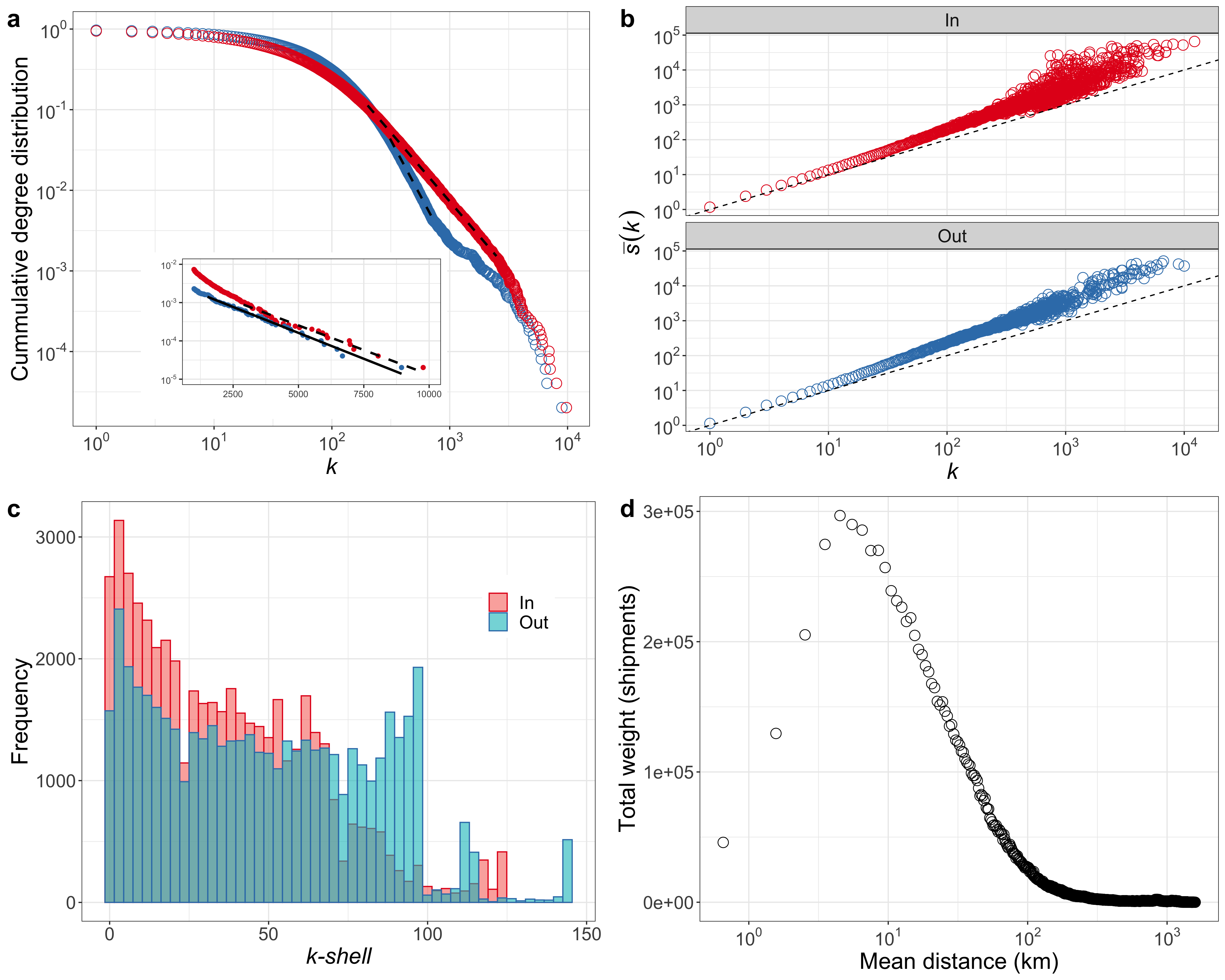}
\caption{\label{fig:02} {\bf a}. Log-Log plot of the complement of the cumulative distribution of in (red) and out (blue) degrees (The inset shows the exponential decay of the tail for both distributions). {\bf b}. Average strength $\bar{s}(k)$ of epiunits with degree $k$ for in/out degree/strength, as indicated in panels. Dashed line indicates $s_i=k_i$ {\bf c}. Frequency of epiunits in each (in - red/out - blue) $k$-shell in the network. {\bf d}. Total added weight of shipments versus the distance traveled for the shipment (in kilometers).}
\end{figure}

Both the in and out degree distributions have a scale-free like behavior for values of $k_i^{in}\in(200,2500)$ for the in-degree, and $k_i^{out}\in(300,750)$ for the out-degree (Figure \ref{fig:02}{\bf a}). For large $k$ ($k^{in}>3000$; $k^{out}>1500$) both distributions show an exponential cut-off (Figure \ref{fig:02}{\bf a} - inset). The degree distribution of the shipments network (in and out) is more similar to a log-normal distribution than to other candidates considered ($\Delta_2^{in}=5457.3$, $\Delta_2^{out}=278.6$) (Supplemental material Section 3). To investigate the relationship between the strength and degree of epiunits, we analyze how these quantities depend on each other (Figure \ref{fig:02}{\bf b}). It has been observed that $\bar{s} \propto k^{\beta}$, with $\beta\geq1$ for single \cite{Barrat3747} and multiplex \cite{menichetti_weighted_2014} networks.  A value of $\beta=1$ shows a linear relationship between $s$ and $k$, meaning that, on average, the weights of the links upon the hubs are similar to the weights of the links of less connected nodes. On the other hand, when $\beta>1$ the weights of the links connected to the hubs are larger than the weights of the links of less connected nodes. The shipments network shows that the strength of the epiunits with degree $k$ increases super-linearly with $k$ for both, the in and out degrees (Figure \ref{fig:02}{\bf b}) with $\beta^{in} = 1.22 \pm 0.01$ and $\beta^{out}=1.21 \pm 0.006$. The super-linear relationship between $s$ and $k$ implies that the strength of epiunits grows at a faster pace than their degree, which denotes that the more shipments an epiunit sends/receives, the more frequent the shipments tend to be sent/received by that epiunit.

The shell decomposition of the shipments network, which has been shown to be predictive of the spreading potential of nodes \cite{kitsak2010}, is shown in Figure \ref{fig:02}{\bf c}. The process of decomposing the shipments network into its $k$-shells, follows the steps: first, remove all epiunits with one connection only (and their links), until no more such epiunits remain, these epiunits are assigned to the $1$-shell. Recursively, we remove all epiunits with degree 2 (or less), these epiunits are in the $2$-shell. Increasing $k$, we repeat this process until all epiunits in the shipments network have been assigned to one of the shells. The $k$-core (consequently, the coreness of an epiunit) is defined as the union of all shells with indices larger or equal to $k$. We observe that the \emph{nucleus} \cite{Carmi11150} of the shipments network (all epiunits in the highest shell) is formed by $0.8\%$ ($0.9\%$) of all epiunits with respect to the in-coreness (out-coreness) of the network. There is a monotonic decrease in the number of epiunits as the $k$-shell increases using the in-coreness (Figure \ref{fig:02}{\bf c}). However, for the out-coreness, the deepest part of the network is composed by a small number of epiunits ($k$-shell$\geq100$).

In the shipments network, exchanges of cattle occur at many scales (Figure \ref{fig:02}{\bf d}). Above a distance of $5$ km, the weight of shipments (frequency of shipments between epiunits) decreases with distance; becoming very rare for shipments longer than $~120$ km. Notably, the weight and the mean distance of shipments is positively correlated for small distances, which may reflect: the inherent spacing of epiunits; a restriction of interaction between close epiunits; lack of data recording at this level; among others.

\begin{figure}[!ht]
\centering
\includegraphics[width=0.47\textwidth]{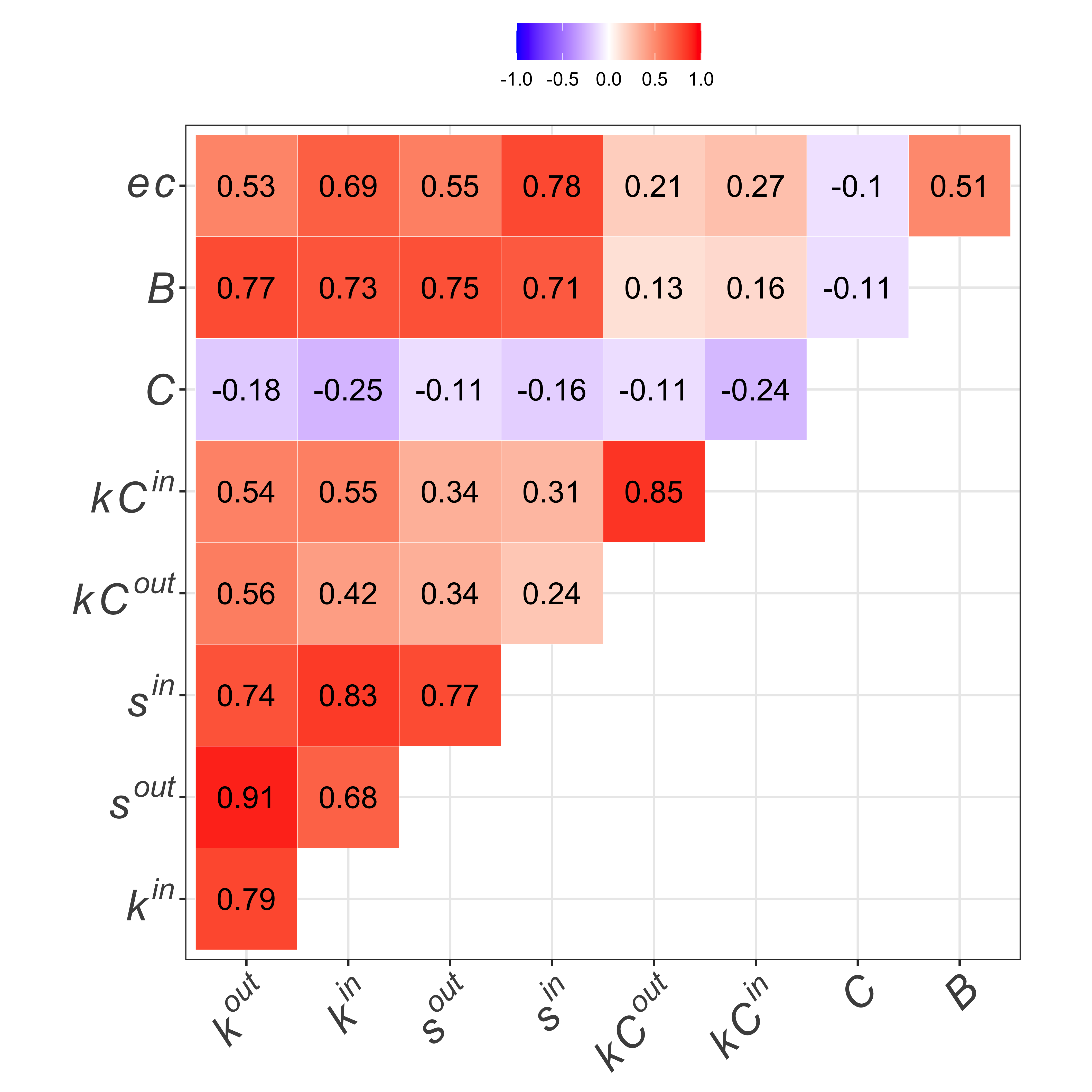}
\includegraphics[width=0.46\textwidth]{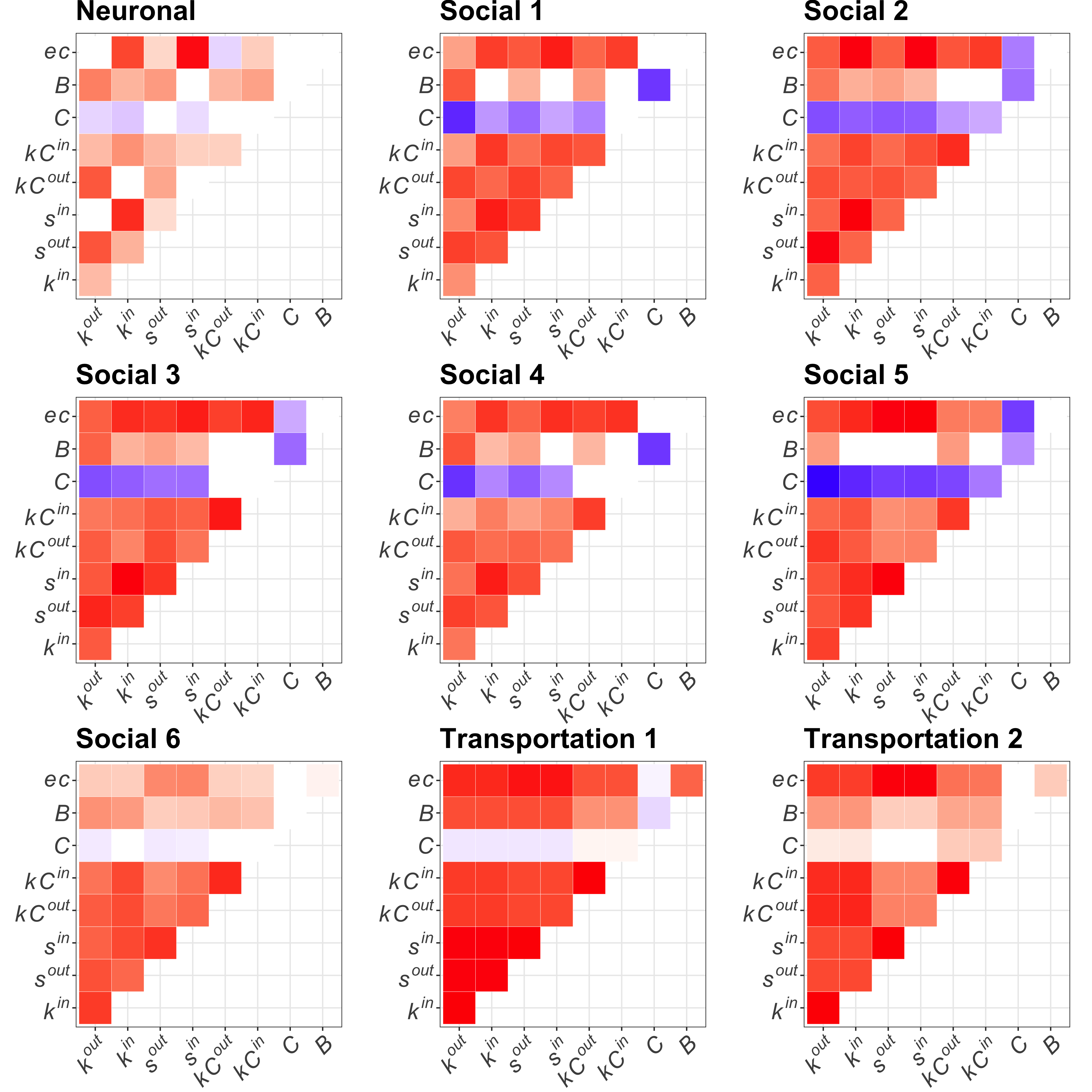}
\caption{\label{fig:04} Correlation matrices between ranking of nodes using different measures. 
ec: eigenvector centrality; B: Betweenness centrality; C: clustering coefficient; $kC^{in}$: in-coreness; $kC^{out}$: out-coreness; $s^{in}$: in-strength; $s^{out}$: out-strength; $k^{in}$: in-degree and $k^{out}$: out-degree. 
Left: correlations between measures for the shipments network. Right: correlations between measures for 9 other networks from different contexts (neuronal, social and transportation) for comparison. (Detailed information about networks on the right can be found in Section 3 of the supplemental material).} 
\end{figure}

The shipments network exhibits high correlations between the in- and out- modes of measures (degree ($k^{in/out}$), strength ($s^{in/out}$) and coreness ($kC^{in/out}$)) (Figure \ref{fig:04}-left). The betweenness centrality ($B$) is highly correlated with degree and strength, implying that epiunits with high movements of cattle serve as bridges that connect different regions. Clustering coefficient ($C$) is negatively correlated with all other measures. Eigenvector centrality ($ec$), a measure that evaluates the quantity and quality of connections of an epiunit, shows a positive correlation with most network measures except $C$; epiunits with the largest degree and strength are not the most central in the network, according to the eigenvector centrality. Epiunits with many shipments are not necessarily those that are best connected in the shipments network. Epiunits located in the nucleus of the network (deep in the network, according to in/out coreness) are not the most central in the shipments network (low correlation with $ec$), or bridges (low correlation with $B$); however, epiunits with high in-coreness tend to be epiunits with high out-coreness as well (Figure \ref{fig:04}-left).

To assess the general features of the shipments network and to associate it with other families of network models, we evaluate the correlations among different local measures of centrality in the shipments network (Figure \ref{fig:04}-left), and compare it qualitatively to other models of networks, coming from other contexts (Figure \ref{fig:04}-right). Qualitatively, the 5 social networks show similar negative correlation between clustering (C) and other node measures. Overall, the patterns of correlation among measures in the shipments network is most similar to Transportation 1 ($0.91$), and Social 2 ($0.85$) and most different from Neuronal ($0.75$).

\subsection{FMD outbreaks in Turkey}

\begin{figure}[!ht]
\centering
\includegraphics[width=\textwidth]{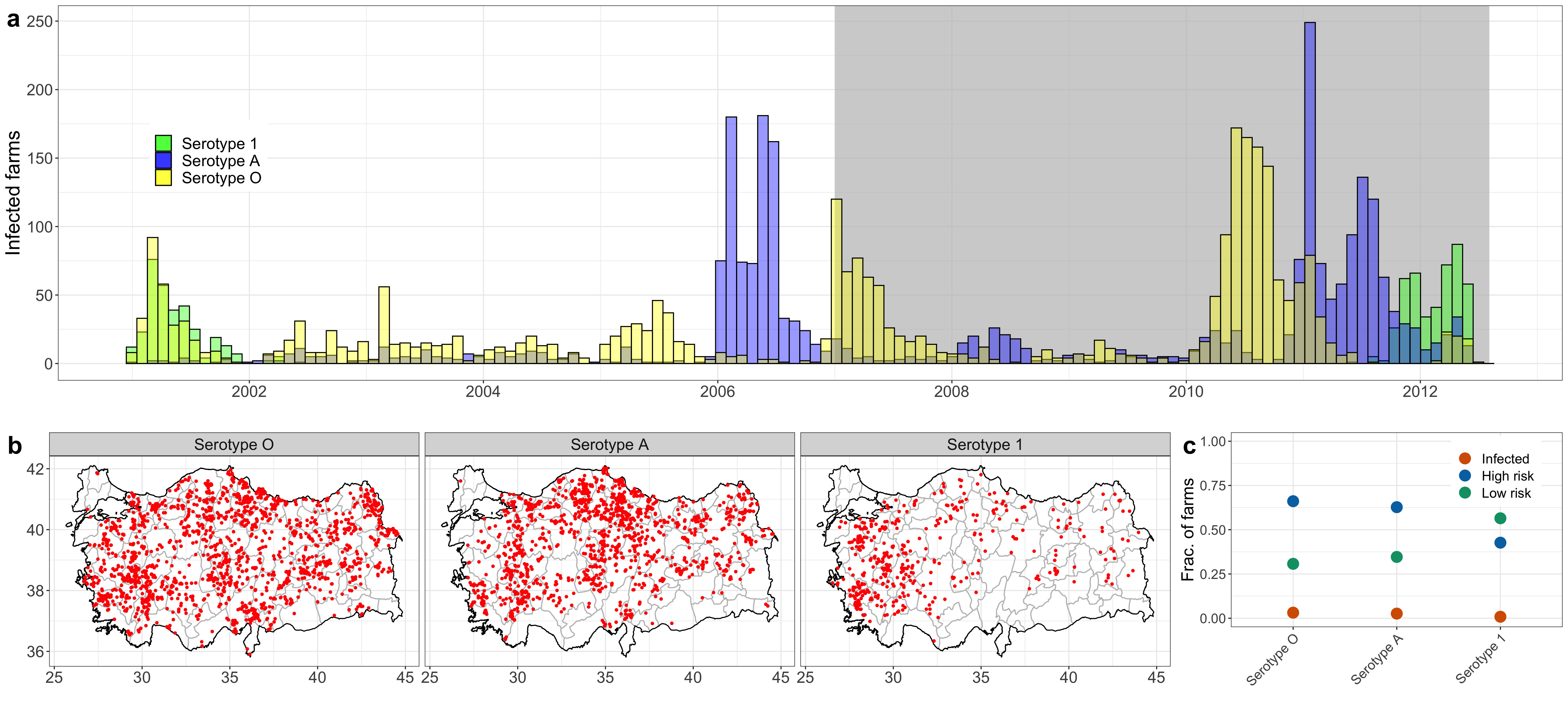}
\caption{\label{fig:outs} {\bf a}. Incidence (accumulated by month) of each strain of FMD in Turkey. The gray region shows the time range where the outbreak data and the shipments data overlap (2007 - 2012). {\bf b}. Map of Turkey showing the location of each of the epiunits (red points) where there occurred an outbreak of FMD. Each panel shows the location of epiunits where occurred at least one outbreak, according to serotype. {\bf c}. Fraction of epiunits in each state (infected (red), High risk (blue) and Low risk (green), for each FMD serotype.}
\end{figure}

From January 2001 to July 2012, serotypes O and A were endemic throughout Turkey, with continuous low prevalence throughout the country (Figure \ref{fig:outs}{\bf b}) and large outbreaks in 2010 and 2011, respectively. The Asia-1 serotype re-emerged Turkey in 2011 (Figure \ref{fig:outs}{\bf a}) and was disproportionately concentrated in the west of the country (Figure \ref{fig:outs}{\bf b}). 

Of all $49,580$ epiunits in the shipments network, $3437$ were involved in at least one outbreak of any of the serotypes between January, 2007 to July, 2012.
Serotype O was detected in $1637$ epiunits, $1356$ epiunits were infected with serotype A, and $444$ epiunits infected with serotype Asia-1. Most epiunits experienced only 1 outbreak of any serotype; though $1$ epiunit had $8$ outbreaks of serotype A, $1$ epiunit had $5$ outbreaks of serotype O, and one epiunit had $3$ outbreaks of serotype Asia-1. Of all epiunits, $19$ experienced outbreaks of all three serotypes.

The fraction of epiunits that were ever infected are $3.2\%$, $2.6\%$ and $0.8\%$ for serotypes O, A and Asia-1, respectively (Figure \ref{fig:outs}{\bf c}). For serotypes A and O (endemic), epiunits that were \emph{low-risk} of infection (at distance 2 or more links) were the $34.6\%$ and $30.7\%$ of the population of epiunits, respectively; while most epiunits were \emph{high-risk} (no outbreak, but an outbreak in a neighboring epiunit; $62.8\%$ and $66.1\%$, respectively). Epiunits infected with serotype Asia-1, which re-emerged in 2011, showed a different pattern from the endemic serotypes; $56.5\%$ epiunits were at low risk, with $42.7\%$ epiunits at high risk of getting infected.

\begin{figure}[!ht]
\centering
\includegraphics[width=\textwidth]{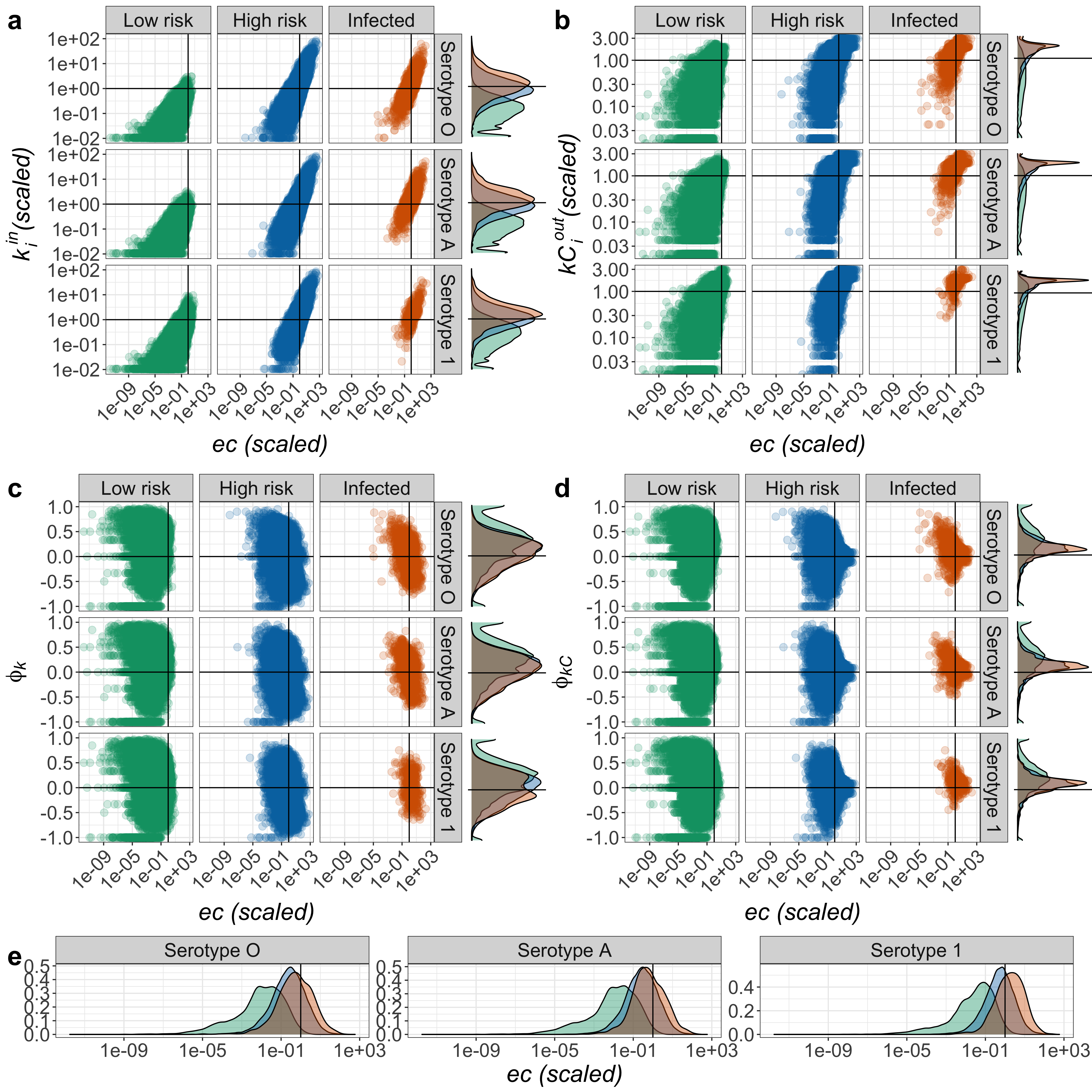}
\caption{\label{fig:07.1} {\bf a}. Correlation plane ($ec$,$k_i^{in}$). Marginal plots correspond to densities of $k_i^{in}$. {\bf b}. Correlation plane ($ec,kC_i^{out}$). Marginal plot corresponds to the densities of $kC_i^{out}$. {\bf c} Correlation plane ($ec$,$\phi_k$). Marginal plots correspond to densities of $\phi_k$. {\bf d} Correlation plane ($ec$,$\phi_{kC}$). Marginal plots correspond to densities of $\phi_{kC}$.
{\bf e}. Density plots for $ec$ for each of the three serotypes. $k_i^{in}$, $kC_i^{out}$ and $ec$ have been re-scaled using their corresponding mean value in the complete network. Horizontal lines show, for $k^{in}$ and $kC_i^{out}$, the location of the mean value and $\phi_k=0$ and $\phi_{kC}=0$ for reference. In all panels colors correspond to low risk (green), high risk (blue), and infected (red) epiunits for each of the three serotypes. Vertical lines show the location of the mean value for the $ec$ in all plots. Different strains of FMD are as labeled.}
\end{figure}

Considering the fraction of centrality accumulated by epiunits (the ratio between the centrality of each epiunit and the total sum of centrality in the network) depending on their state, we observe that the epiunits infected by the endemic serotypes (O and A)  accumulated $\approx10\%$ of the total sum of normalized centrality of the shipments network regardless of the measure. By contrast, epiunits infected with the Asia-1 serotype ($0.8\%$ of all epiunits) comprise approximately $1\%$ of total sum of normalized centrality. Low-risk epiunits are under-represented with respect to network centrality. For serotypes O and A, low-risk epiunits accounted for $\approx30\%$ of all epiunits; these epiunits reflect $\approx0.1\%$ of eigenvector centrality and $\approx10\%$ out-coreness centrality, implying that these epiunits are disproportionately on the edge of the shipments network. Epiunits that were low-risk with respect to the Asia-1 serotype were also disproportionately non-central in the shipments network, though less so than the endemic serotypes (Section 5 in supplemental material). 

For all serotypes, there is a positive correlation between in-degree ($k^{in}_i$) and eigenvector centrality ($ec$; Figure \ref{fig:07.1}{\bf a}). Low-risk epiunits had disproportionately lower in-degree ($k^{in}_i$) and eigenvector centrality ($ec$) than infected epiunits, with nearly all low-risk epiunits below the mean value for both measures. Infected epiunits had a slightly higher mean $k^{in}_i$, but comparable $ec$, than high-risk epiunits for both of the endemic serotypes, O and A (Figure \ref{fig:07.1} marginal plots). Epiunits infected with the Asia-1 serotype tend to have higher eigenvector centrality ($ec$) than the mean for the network. When considering out-coreness ($kC_i^{out}$; Figure \ref{fig:07.1}{\bf b}) the positive correlation between $kC_i^{out}$ remains (weakly) for low-risk epiunits. High-risk epiunits span over all ranges of $kC_i^{out}$, with a strong concentration of epiunits with out-coreness larger than the mean. Infected epiunits are mainly those with high values of out-coreness, making epiunits in the nucleus of the network more predictive of infection than degree or eigenvector centrality.

For all serotypes, there is a weak negative correlation between transmission flux of degree ($\phi_k$) and eigenvector centrality ($ec$; Figure \ref{fig:07.1}{\bf c}), implying that epiunits with lower eigenvector centrality ($ec$) tend to be net sources of shipments ($\phi_k$ > 0). Compared to high-risk and infected epiunits, low-risk epiunits are disproportionately sources of shipments ($\phi_k$ > 0), and many are exclusively sources ($\phi_k$ =1). For the transmission flux of coreness ($\phi_{kC}$), the behavior is similar to $\phi_k$, but with a more narrow concentration of epiunits (specially high-risk and infected epiunits) as mostly net sources of shipments (Correlation plots for other network measures used are presented in Section 6 of the supplemental material). Even when the eigenvector centrality ($ec$) and the out-coreness ($kC^{out}$) are not well correlated (Figure \ref{fig:04}-left), epiunits that get infected have high values of $ec$ and can be found mostly in the nucleus of the network (Figure \ref{fig:07.1} {\bf b}).

We compared the properties of epiunits during the large outbreaks of serotypes O, A, and Asia-1 between 2010-2011. These epiunits had similar behavior in in-degree ($k^{in}_i$), eigenvector centrality ($ec$) and transmission flux of degree ($\phi_k$) as those during the non-outbreak period from 2009-2010 (supplemental material, Section 7). Similarly, the properties at the start of a large outbreak %(first 25\% of infected nodes)
(epiunits involved in the fast initial growth of the outbreak, i.e. while the first and second rates of change of the curve remained positive) were similar to those in the rest of the outbreak (supplemental material, Section 7).

Lastly, after removing all epiunits ever infected from the shipment network, along with their connections, the GSCC of the \emph{residual} network is $0.96$, which means that $96\%$ of epiunits in the residual network can be reached from any other epiunit through directed connections, which is evidence of the stability of the shipments network in Turkey. Unfortunately, this stability leaves the transportation network vulnerable to a subsequent introduction of FMD.

\section{Discussion}

In the present paper we used a large and detailed database of shipments of cattle between epiunits in Turkey, to build and characterize an aggregated directed-weighted network. In addition, with data corresponding to outbreaks of FMD in Turkey, we study the correlations between network-based ``importance'' of epiunits and their reported FMD state (infected, high-risk and low-risk epiunits). The shipments network shows a combination of features that influence the way in which a disease may spread
\cite{watts1998,newman2003,newman2002,schwartz2002,Eames13330,kitsak2010,makarov2018,newman2002a}.

The shipments network has strong evidence of small-worldness. High values of the clustering coefficient indicate that epiunits are prone to the formation of closed triangles of shipments, which has not been observed for previous shipments network studies \cite{kiss2006}. However, our value of clustering is an upper bound, which would decrease when introducing directions in the links. High clustering implies that epiunits would tend to spread a disease more rapidly within clusters, though with smaller outbreaks (total number of infected epiunits) than in a random network, and a reduction of the epidemic threshold \cite{newman2003}. The ``small'' attribute of the shipments network is given by the shortest path length ($L=2.86$), comparable to a random equivalent network; an indication of high likelihood of infection of an epiunit, provided that a close epiunit is already infected. Despite the redundancy in local spread implied by high clustering, the small diameter of the network ($D=20$), along with a GSCC comparable with the complete network ($97\%$), indicates that a spreading pathogen could traverse and infect the shipments network using directed links in very few infectious generations. Both values of $L$ and $D$ are lower bounds of these measures in the shipment network, and will increase when considering the weights of links in their calculation; however, these values are still expected to be low (similar to its random equivalent network), due to the strong cohesiveness of the shipments network (GSCC$=97\%$). Consequently, the largest eigenvalue of the adjacency matrix of the shipment network ($\lambda_1$) is large, and considerably higher than a random equivalent network implying that the shipments network is inherently susceptible to large outbreaks, since it has been shown that the epidemic threshold of a network depends on the reciprocal of $\lambda_1$ \cite{wang2003}.

The shipments network shows a small tendency to a dissasortative mixing of degrees ($\rho$ small but negative) which indicates that epiunits send/receive shipments of cattle to/from other epiunits of different degrees, as seen, for instance, for technological \cite{newman2002a}, spatial \cite{makarov2018} and other cattle transportation networks \cite{kiss2006}. The presence of long range connections, where local \emph{hubs} (highly connected epiunits) can connect with other hubs, also promotes connections with peripheral epiunits, decreasing the assortativity in spatially restricted networks \cite{makarov2018}. The high degree of modularity ($Q$) in the Turkey shipments network implies a strong local connectivity, where the number of defined communities and its correlations with spatial distribution needs a more in-depth consideration and study.

The frequency of shipments (weights of connections) is low among very near (positively correlated) and very far epiunits (Figure \ref{fig:02}{\bf d}). The former may reflect the underlying spatial distribution of the epiunits themselves, evidenced by a strong community structure (high modularity); this structure may be driven by potential restrictions of cattle exchange between very closely located epiunits, perhaps arising from competition between them or from failure to record local shipments. The latter suggests that there may be a cost to shipments among distant locations. This local community structure is analogous to that seen in metapopulation dynamics \cite{cross2007} and has previously been shown to impact the dynamics of epidemic spread \cite{apolloni2014,Sah4165}, with separate transmission thresholds for spread within modules and global outbreaks \cite{gross2020}. Similarly, high modularity suggests the potential benefit of locally targeted surveillance and control strategies \cite{gross2020}. Additionally, the correlation of the rankings of epiunits in our network (Figure \ref{fig:04}) shows that central epiunits (high eigenvector centrality) are not necessarily those with more connections to/from it (in/out degree). Several epiunits with influential connections (high eigenvector centrality) serve as \emph{bridges} of densely connected epiunits (high betweenness centrality) (Figure \ref{fig:04}); i.e. are connections between modules. These attributes suggest that network-based active surveillance and control strategies on the shipments network, should be managed locally, identifying central epiunits (potential sentinel epiunits), with fluxes (of degree or coreness) close to zero, to prevent the infection of hubs and rapid within community infection. Bridges of transmission, located between communities, with a high betweenness centrality, could be targeted to prevent transmission between communities.

When compared to other models of networks, the correlation in the ranking of epiunits in the Turkey shipments network is most similar to social and transportation networks (Figure \ref{fig:04}).
The low reciprocity implies a net directional movement of cattle. These attributes, along with a preference for shipments between epiunits at a typical distance (Figure \ref{fig:04}{\bf d}) and the exponential cut-off in the degree distribution (Figure \ref{fig:04}{\bf a} - inset), imply a spatial-like distribution of epiunits along Turkey's geography, where epiunits appear to be clustered in each of the provinces ($81$) due to geographical and/or other restrictions.

In contrast to shipments networks characterized elsewhere \cite{kao2006,lentz2016,kiss2006,rowland2007,MOHR20188,kaluza2010}, the Turkey cattle shipments network does not display scale-free behavior in all ranges of its degree distribution. Instead, only intermediate values of the in/out degree show scale-free behavior, with an exponential cut-off for large in/out degree (Figure \ref{fig:02}{\bf a}). The in/out degree distributions of the shipments network of Turkey are more similar to a log-normal than other distributions tested (Supplemental material, Section 3). Moreover, the positive and high correlation between in-degree ($k^{in}$) and out-degree ($k^{out}$) (Figure \ref{fig:04}) implies that a disease spreading in the shipments network is boosted by the existence of epiunits that could get infected from many sources (large $k^{in}$) and that also could infect many other epiunits (large $k^{out}$).

Though FMD outbreaks between 2007-2012 were broadly distributed across the country (Figure \ref{fig:outs}), there were distinct relationships between node-level network properties and the likelihood that an epiunit would experience an outbreak. Notably, epiunits that never recorded an outbreak had disproportionately lower in-degree, lower out-coreness, lower eigenvector centrality, and were more likely to be sources than sinks of shipments compared to epiunits that had outbreaks or were at high risk (one link away from an outbreak). Epiunits infected with the endemic strains, A and O, and high-risk epiunits had similar distribution of node characteristics. Together, this suggests that while it is difficult to distinguish epiunits that will experience outbreaks from those that will be nearby, node-level network metrics are predictive of epiunits that are unlikely to be infected or exposed. The distribution of the re-emerging serotype, Asia-1, is intriguing in that the infected nodes had disproportionately high eigenvector centrality. That both central and peripheral epiunits were frequently infected might be a consequence of the spatial distribution of epiunits, with both central hubs and low degree bridge epiunits. These observations warrants further investigation. 

Theory predicts that both network-level and node-level measures should be correlated with infection risk. Here we have described a complex, real-world network that conforms to some patterns of classically studied models, with properties of small-world networks and spatially constrained edge probabilities. We have shown node-level properties are only weakly correlated with the likelihood of infection with an endemic disease (i.e. weak sensitivity), but these same properties are highly predictive of nodes that were never exposed (i.e. strong specificity). 

Our results show that there exist intricacies which need to be considered when implementing surveillance strategies in a population. Even when complete knowledge of the contact structure of a population is granted, along with information about an epidemic/endemic disease on the network, there are complex interactions between these two components that may obscure our predictive power. Considering high ranked nodes in a network \cite{herrera2016,bai2017,schirdewahn2021,kitsak2010,christley2005,christakis2010} is an important starting point; however, as shown here, some patterns will only be revealed by evaluating the empirical distribution of infection on these networks relative to theoretical predictions. Lastly, the inclusion of explicit temporal variability (changes in the interactions between epiunits, and network global/local features) \cite{holme2012}, adaptive behavior (change in shipments patterns of epiunits, due to change in the ``state'' of epiunits (i.e. infected or vaccinated), which would eventually change the shipment patterns) \cite{gross_epidemic_2006} and other details (if plausible) in the interactions of epiunits will be exciting avenues for future research.

\bibliographystyle{numeric}

%\bibliography{FMD-project}

\end{document}